\documentclass[sigconf]{acmart}

\usepackage{url}  
\usepackage{amsmath}
\usepackage{graphicx} 
\usepackage{graphicx}  
\usepackage{xcolor}
\usepackage{hyperref}

\usepackage{multirow}
\usepackage{algpseudocode}\usepackage{algorithm}
\usepackage{bbm}
\usepackage{comment}

\AtBeginDocument{%
  \providecommand\BibTeX{{%
    \normalfont B\kern-0.5em{\scshape i\kern-0.25em b}\kern-0.8em\TeX}}}

\setcopyright{iw3c2w3}

\begin{document}

\copyrightyear{2021}
\acmYear{2021} 
\acmConference[WWW '21]{Proceedings of the Web Conference 2021}{April 19--23, 2021}{Ljubljana, Slovenia} 
\acmBooktitle{Proceedings of the Web Conference 2021 (WWW '21), April 19--23, 2021, Ljubljana, Slovenia}
\acmPrice{}
\acmDOI{10.1145/3442381.3449845}
\acmISBN{978-1-4503-8312-7/21/04}

\title{
Causal Network Motifs: Identifying Heterogeneous Spillover Effects  in A/B Tests}

\author{Yuan Yuan}
\affiliation{
\institution{Massachusetts Institute of Technology} 
\city{Cambridge, MA} \country{USA}}
\email{yuan2@mit.edu}

\author{Kristen M. Altenburger}
\affiliation{\institution{Facebook Core Data Science} 
\city{Menlo Park, CA} 
\country{USA}}
\email{kaltenburger@fb.com}

\author{Farshad Kooti}
\affiliation{\institution{Facebook Core Data Science} 
\city{Menlo Park, CA} \country{USA}}
\email{farshadkt@fb.com}

\begin{abstract}
Randomized experiments, or ``A/B'' tests, remain the gold standard for evaluating the causal effect of a policy intervention or product change. However, experimental settings, such as social networks, where users are interacting and influencing one another, may violate conventional assumptions of no interference for credible causal inference. Existing solutions to the network setting include accounting for the fraction or count of treated neighbors in a user's network, yet most current methods do not account for the local network structure beyond simply counting the number of neighbors.  Our study provides an approach that accounts for both the local structure in a user's social network via motifs as well as the treatment assignment conditions of neighbors. We propose a two-part approach. We first introduce and employ ``causal network motifs'', which are network motifs that characterize the assignment conditions in local ego networks; and then we propose a tree-based algorithm for identifying different network interference conditions and estimating their average potential outcomes. Our approach can account for social network theories, such as structural diversity and echo chambers, and also can help specify network interference conditions that are suitable to each experiment. 
We test our method on a synthetic network setting and on a real-world experiment on a large-scale network, which highlight how accounting for local structures can better account for different interference patterns in networks.

\end{abstract}

\maketitle

\section{Introduction}

Randomized control trials, or ``A/B tests'', have been crucial to understanding the impact of an intervention, such as a new policy~\cite{ho2017does,handan2020feasible}, product intervention~\cite{aral2011creating,bakshy2012social}, or medical treatment~\cite{antman1992comparison,loucks2003luteinizing}.
Randomized control trials can estimate the causal effect of a treatment intervention by ensuring that treatment and control assignments are independent of other variables. Increasingly, causal inference methods have had to adapt to modern A/B test settings where there are high-dimensional features~\cite{savje2017generalized,roberts2018adjusting,egami2018make,eckles2020bias}, computational and algorithmic considerations~\cite{wong2020computational}, and network interference concerns~\cite{aral2016networked,eckles2016design,saveski2017detecting}.

For addressing interference, traditional causal inference methods rely on a critical assumption called the ``stable unit treatment value assumption'' (SUTVA)~\cite{fisher1937design,splawa1990application}. SUTVA is in fact a strong assumption, requiring that a unit's outcome is only affected by its own assignment conditions, regardless of the assignment conditions of all other observations.
However, this can be an unrealistic assumption in many settings, including social networks, where user's are influenced by one another ~\cite{pouget2019testing,bowers2013reasoning,athey2018exact}. We refer to this as \textit{network interference} \cite{aral2016networked}.

An increasing number of approaches are aimed at dealing with network interference~\cite{rosenbaum2007interference,tchetgen2012causal,ugander2013graph,kohavi2013online,xu2015infrastructure,aronow2017estimating,basse2018analyzing,saveski2017detecting,imai2020causal}. Existing methods for addressing networked interference can be categorized into two main strategies. The first one is to improve the random assignment strategy. Cluster random assignment treats observations at the level in which observations have strong interdependence (e.g., assigning on the class or city level)~\cite{rosenbaum2007interference,gerber2012field,basse2018analyzing}. Graph cluster randomization is a special case for social networks~\cite{ugander2013graph,eckles2016design}. It first runs a graph clustering algorithm and then randomizes treatment assignment on the graph cluster level. Recent work proposes random clustering instead of a fixed clustering~\cite{ugander2020randomized}.
Another type of approach, including ours, relaxes SUTVA by allowing the potential outcome to be defined as a function of the assignment conditions of the ego observation and its ($n$-hop) neighbors~\cite{vanderweele2008ignorability,leung2020treatment}. However, few studies have utilized the network structure in the $n$-hop neighbors (i.e. local network structure) to further characterize different interference conditions. 

\begin{figure}[t!]
    \centering
    \includegraphics[width=\linewidth]{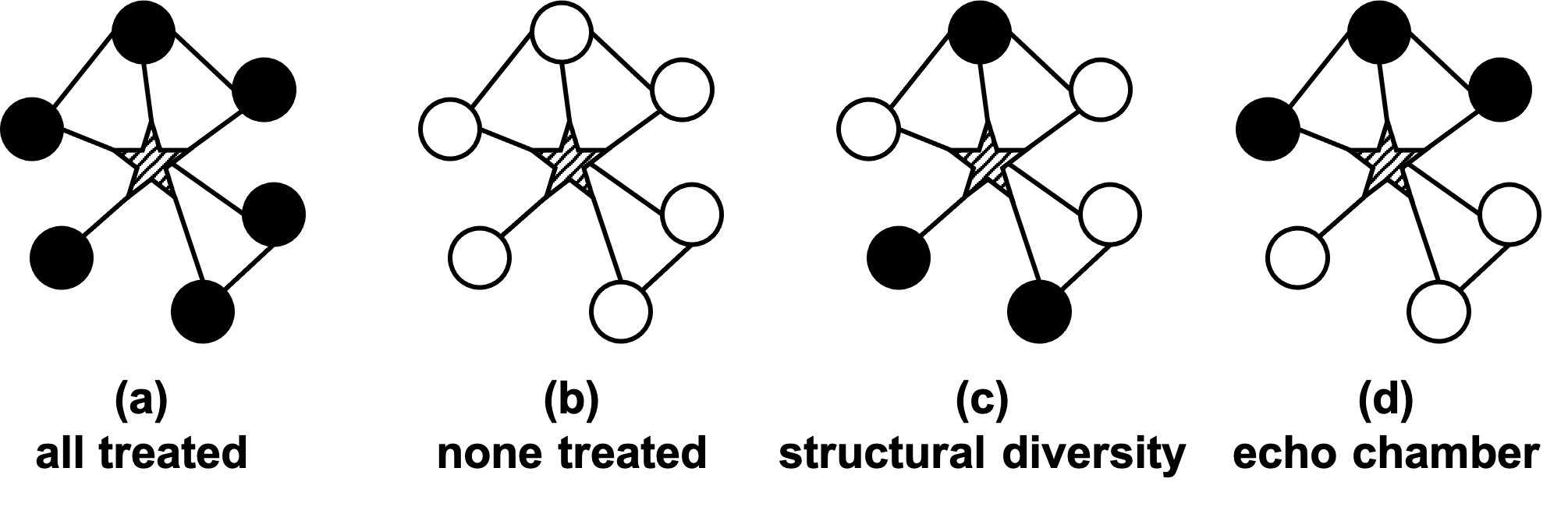}
    \caption{\small{\textbf{Examples of network interference conditions across different local network structures.} The star indicates a user and a circle represents a user's friends. Solid circles indicate that a friend is in treatment and hollow circles indicate a friend is in control. For stars, the shaded indicates that it could be treated or control.}}
    \label{fig:spillover}
\end{figure}

From the empirical end, many social network studies have highlighted why local structures should be considered to address network interference. For example, the structural diversity hypothesis \cite{ugander2012structural,su2020experimental} claims the likelihood of product adoption is largely dependent on the degree to which a unit's neighbors who have adopted are disconnected. By contrast, complex contagion theory \cite{centola2007complex,centola2010spread} or the ``echo chamber'' effect \cite{bakshy2015exposure,flaxman2016filter}, suggest an individual is most likely to adopt a behavior when she is clustered in the network of multiple neighbors who have adopted this behavior. Figure~\ref{fig:spillover} illustrates four examples of network interference that could be captured by the local networks structure and treatment assignment. The first two conditions are simply the cases where all neighbors are treated or non-treated, followed by the important network interference conditions suggested by structural diversity and complex contagion, respectively. In the case of structural diversity and echo chamber settings, the ego node in (c) and (d) has 1/2 neighbors treated but exhibit very different local structures and the ego's outcome may be  different in these settings; and we do not know which one is the dominant factor that drives most of the variance in the outcome.

Our study provides a tool for experimenters and practitioners to account for important network interference conditions without necessarily specifying a particular dominant social science theory. We rely on network motifs~\cite{milo2002network,alon2007network} (or graphlets~\cite{sarajlic2016graphlet}) to characterize both the local structure and treatment assignment conditions to account for the different types of (network interference) exposure conditions. We can think of network motifs as ``label-independent structure'' since the features do not depend on treatment labels and network motifs with treatment assignment conditions as ``label-dependent structure'' since they depend on labels~\cite{gallagher2008leveraging}. We refer to network motifs with treatment assignment labels as \textit{causal network motifs}. Prior work has  considered label-dependent network motifs for use in protein function prediction applications~\cite{chen2007labeling} and has developed computationally tractable ways for detecting labeled motifs~\cite{ribeiro2014discovering}. Another example is ~\cite{arpino2015implementing}, which uses network features such as transitivity in a propensity score matching framework, with a focus on observational data. Our proposal is distinct in that we focus on experimental settings where causal network motifs can quantify the causal impact of different interference conditions.

In this paper, we provide an approach to automatically categorize network interference conditions based on local network neighborhood structures. We develop our approach based on the framework proposed by \cite{aronow2017estimating}, where the authors regard different network interference conditions\footnote{ In this paper, we use exposure conditions and network interference conditions interchangeably.} as unique treatments.
However, it is unclear how experimenters or practitioners should a priori define different network interference conditions that are suitable for specific data and experiment. Moreover,  manually determining different network interference conditions may make more severe the issue of a large number of researcher degrees of freedom.

The following sections will describe our two-step solution to identify different network interference conditions. First, we construct network motifs with assignment labels, referred to as \textit{causal network motifs}, to characterize complex network interference conditions. In our study, we not only characterize the network structure by the number of different network motifs, but also characterize the assignment conditions in the local network neighborhood by the causal network motifs. Second, using these causal network motif features, we develop a tree-based algorithm to cluster different exposure conditions. Each leaf or terminal node in the tree corresponds to a network interference condition and can be used to estimate the average potential outcome given that condition. We cannot directly apply conventional decision trees or other machine learning algorithms, since we need to adjust those algorithms to address issues, such as selection bias and positivity as will be explained~\cite{westreich2010invited,chernozhukov2017double,chin2019regression}. Moreover, as a common goal in causal inference, we aim to estimate the average potential outcome given an interference condition to quantify causal impacts, rather than predicting the potential outcome and the treatment effect for every observation \cite{rubin2005causal}.

The rest of the paper is organized as follows. 
Section~\ref{sec:setup} provides preliminaries of causal inference, the potential outcomes framework, and why the SUTVA may fail in network experiments. Then we illustrate our two-part approach: Section~\ref{sec:motif} presents how we construct the ``causal network motifs,'' which characterize the treatment assignment conditions of each observation as well as their network neighborhood; and Section~\ref{sec:tree} discusses our tree-based algorithm that regards the ``causal network motifs'' as input and then labels each observation as a network interference condition. Section~\ref{sec:experiment} empirically shows the validity of our approach using a synthetic experiment and a real-world experiment. Section~\ref{sec:discussion} concludes.

\section{Causal Inference Setup}

\label{sec:setup}

Let $i$ (or $j$) index individuals in a social network. $\mathcal{U}$ is the set of individuals in the population. $\mathcal{N}_i$ be the neighbor set of $i$, and this can be extended beyond immediate neighbors.
$Y_i$ is the observed potential outcome. $Z_i \in \{0, 1\}$ denotes the random treatment assignment for $i$. Whenever applicable, we use upper-case letters to represent variables or vectors that could be intervened by experimenters or affected by the intervention; and we use lower-case letters for other variables that are not affected by the intervention (such as demographics). Sets are in calligraphy.

\subsubsection*{Potential outcomes framework for network interference}

We start from the potential outcomes framework~\cite{rubin2005causal}:
The potential outcomes framework defines that a unit's outcome is a function of assignment conditions.
\begin{equation}
    Y_i  = y_i(Z_i).
\end{equation}
\noindent $y_i(1)$ and $y_i(0)$ are called potential outcomes and we only ever observe one of them. This function implies no interference, meaning that the potential outcome $Y_i$ does not depend on the treatment assignment of other users. However, this is often an unrealistic assumption in network settings, since the unit's outcome can be dependent on the treatment assignment conditions of neighbors or even any other unit, through, for example, social contagion. 

To overcome this unrealistic assumption, our approach is based on ~\cite{aronow2017estimating} which introduces \textit{exposure mapping}. An exposure mapping is defined as a function that maps the treatment assignment vector to a value; the fraction of treated friends \cite{ugander2013graph} is an example of a simple exposure mapping. Each exposure mapping corresponds to a condition that considers both treatment assignment conditions of both the ego node and her ($n$-hop) neighbor. Under this framework, researchers or practitioners usually need to define a priori a set of \textit{network interference conditions} (also known as \textit{exposure conditions}). We denote this set by $\mathcal{D}$. Then in this framework, the outcome variable is expressed by
\begin{equation}
Y_i = \sum_{d \in \mathcal{D}} y_i(d) \mathbbm{1} [D_i=d].
\label{eq:Z2}
\end{equation}
\noindent Each $d$ corresponds to a pre-specified exposure condition (interference condition). $D_i$ is a random variable indicating $i$'s exposure condition. 
Analyzing different exposure conditions can offer insights into how observations react to different network interference settings, such as comparing (with selection bias corrected) users in control with treated friends to users in control with only control friends.  

Let us introduce three examples.  First,
SUTVA is a special case under this exposure mapping framework (where $\mathcal{D} = \{ego\_treated,$ $ego\_control\}$).
The second example is the existence of direct and indirect effects, as introduced in \cite{aronow2017estimating}: that $\mathcal{D}=\left\lbrace {no\_exposure}, \right.$ ${direct\_exposure},$ ${indirect\_exposure}$, ${direct}$ $+$ $ {indirect}$ ${\_exposure} \}$. Direct effect means the ego or user is treated and indirect effect means any neighbor is treated; thus we have four conditions in total. In our study, we use direct effect to refer to the effect of changing $Z_i$ from 0 to 1 on the outcome variable; and indirect effect refers to any consequence that result from the assignment conditions besides ego node $i$, i.e. $Z_{-i}$.\footnote{Indirect effects will be further specified by $\mathbf{X}_i$ in later sections.}
The third example, the $k$ (or $q$-fractional) neighborhood conditions in \cite{ugander2013graph} fits this framework, with ($2 \times 2$) four different exposure conditions --- whether the ego node is treated $\times$ whether more than  $k$ (or $q$-fractional) of her neighbors are assigned to the same treatment condition with the ego node.

\subsubsection*{Distinguishing between correlation and causation}

The indirect effect may contain a large degree of heterogeneity based on how many ($n$-hop) neighbors are treated and how they are connected. Some of these conditions are theorized in \cite{centola2007complex, ugander2012structural} while other cannot. It is usually challenging to distinguish between what is the causal impact of certain exposure conditions, versus what is usually confounded by  ego's local network structure (for example, users with more friends tend to have their friends less clustered) \cite{su2020experimental}. In other words, selection bias may result from simply taking average over observations within a certain interference condition. For example, imagine that we want to quantify the impact of an indirect effect, so simply taking the average over observations who have at least one treated friend versus taking average over observations with no treated friends. This average may be biased towards observations who have more friends. We illustrate this selection bias issue in Figure~\ref{fig:selection}.

Mathematically, our goal is to estimate the average potential outcome 
\begin{equation}
\bar{y}(d) = \frac{1}{|\mathcal{U}|} \sum_i y_i(d), 
\text{ for all }d \in \mathcal{D}. \end{equation}
This is one of the core goals proposed in \cite{aronow2017estimating}. 
To correct the aforementioned selection bias, we employ inverse probability weighting, such as Horvitz–Thompson estimator and Hajek estimator \cite{seaman2013review} 
Although the Horvitz–Thompson estimator is an unbiased estimator for the average potential outcome, it empirically has unaffordably high variance. Therefore, following~\cite{eckles2016design,aronow2017estimating}, 
we use the Hajek estimator to estimate the average potential outcome for exposure condition $d$ since 
its small bias can usually be ignored in a large sample:

\begin{equation}
\hat{\bar{y}}_{Hajek} (d) = \frac{\sum_{i \in \mathcal{U}}{ \frac{1_i (d) y_i(d)}{\pi_i (d)}}}{\sum_{i \in \mathcal{U}}{ \frac{1_i (d)}{\pi_i (d)}}} =  \frac{\sum_{i \in \mathcal{U}, D_i=d}{ \frac{ y_i(d)}{\pi_i (d)}}}{\sum_{i \in \mathcal{U}, D_i=d}{ \frac{1}{\pi_i (d)}}} ,
\label{eq:hajek}
\end{equation}

\noindent where we denote $\pi_i(d) = \mathbbm{P} [D_i = d]$ and $1_i(d) = \mathbbm{1} [D_i = d]$. $\pi_i(d)$ is the \textit{inclusion probability} of the exposure condition of $i$ being $d$, and is a generalization of propensity scores~\cite{rosenbaum1983central}. It can also be understood as a weighted average over observations with $1_i(d)=1$, where the weight is $1/\pi_i(d)$. 
Therefore, it can be estimated through weighted linear regressions: the coefficient for the constant is the Hajek estimator. Its variance can be estimated via Taylor linearization \cite{sarndal1992model}. 

The probability $\pi_i(d)$ is often challenging to compute analytically. We therefore use Monte Carlo for sufficiently large replicates to approximate $\pi_i(d)$. 
Specifically, we re-run the treatment assignment procedure for $R$ replicates to obtain the empirical distribution of $(Z_i, \mathbf{X_i})$.\footnote{It is challenging to derive this distribution analytically, especially when the random assignment is deployed on the individual level.  } 
Therefore, we can derive the estimated inclusion probability $\hat{\pi}_i(d)$ for any exposure condition $d$. 
For example, we can let $\hat{\pi}_i(d)=\frac{\sum_r{ \mathbbm{1}[D_i^{(r)}=d]}+1}{R+1}$; and we substitute the $\pi_i(d)$  in Equation~\ref{eq:hajek} with $\hat{\pi}_i(d)$. The relative bias $\hat{\bar{y}}_{Hajek}(d)$  to $\bar{y}(d)$ diminishes exponentially as $R$ increases. The details have been discussed in \cite{aronow2017estimating}.

While the exposure mapping framework introduces a tractable way to take into account network interference, it is an open question how we should then account for the treatment assignment conditions of the $n$-hop neighbors and their network connections in Figure~\ref{fig:spillover}. 
We hope to provide an approach that generates exposure conditions that are suitable for a given experiment and dataset, and that avoids manually defining exposure conditions a priori.
Our study provides a two-step solution to automatically identify different exposure conditions while overcoming selection bias concerns, as will be explained in more detail in the next sections. First, for an A/B test on a network, we construct network motif features with treatment assignment conditions to provide a fine-grained characterization of the local network structure and potential interference conditions.
Second, using the network motif characterization as input, we develop a tree-based algorithm to perform clustering and define the set $\mathcal{D}$ rather than allowing practitioners to explore that.

\section{Causal Network Motifs}
\label{sec:motif}

\begin{figure}
    \centering
    \includegraphics[width=0.7\linewidth]{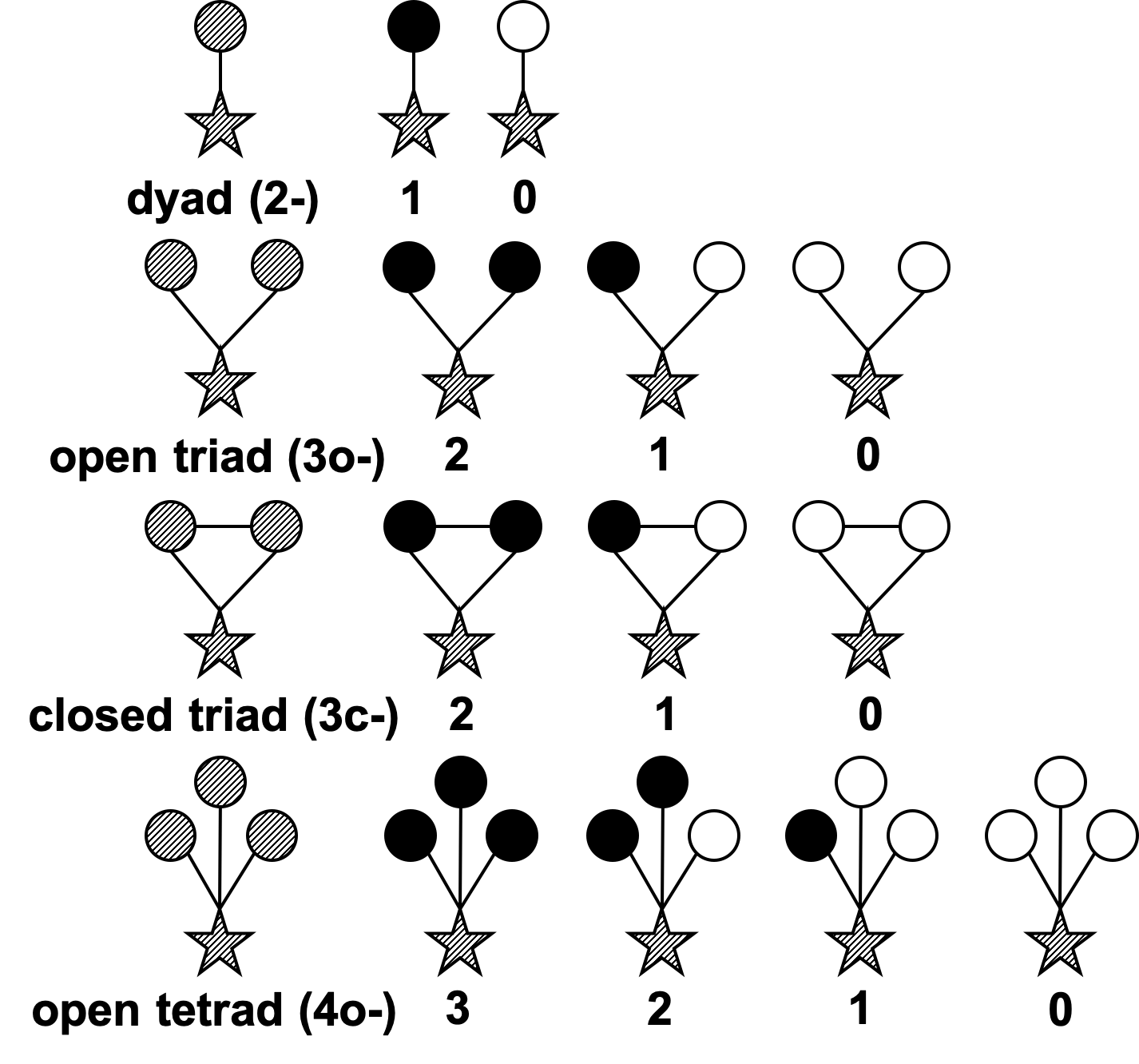}
    \caption{ 
    \small{\textbf{Examples of causal network motifs.} Stars represent egos and circles represent alters. Solid indicates the node being treated, hollow indicates control, and shaded indicates that it could be treated or control. The first patterns in each row are conventional network motifs without assignment conditions, or just called \textit{network motifs}, followed by corresponding network motifs. Our interference vector is constructed by dividing the count of a causal network motif by the count of the corresponding causal network motif.  The labels below each network motif indicate the naming: for example, an open triad where one neighbor is treated is named 3o-1. 
    }}
    \label{fig:motif_example}
\end{figure}

Network motifs are a way to characterize all patterns of smaller network features among a set of nodes~\cite{milo2002network}. We introduce ``causal network motifs,'' which differ from conventional network motifs in two primary aspects. First, we focus on (1-hop) ego networks that include the ego node, with the methods generalizing to higher $n$-hop ego networks for $n$>1. Second, we consider the treatment assignment conditions of the user and their $n$-hop connections. We use the terminology ``network motifs'' to refer to conventional motifs without treatment assignment labels (or assignment conditions) and causal network motifs to refer to ones with assignment conditions. 
Examples of network motifs are illustrated in Figure~\ref{fig:motif_example}. We use these counts on an $n$-hop ego network to characterize the exposure condition of each observation.

Experimenters and practitioners need to determine a priori the region of ego networks (the $n$ for $n$-hop ego networks which means the path length between an ego node and another node is no greater than $n$) and the network motifs that matter. 
For example, Figure~\ref{fig:motif_example} specifies 1-hop ego network and uses the assignment conditions of dyads, triads, and open tetrads as features in the interference vector.

Using causal network motifs implies the following two assumptions:
\begin{enumerate}
\item \textbf{($n$-hop ego networks)} We assume that an ego node's outcome can only be affected by its own assignment condition and its ($n$-hop) neighbors' assignment conditions. This is a common assumption in the prior network interference literature, and is sometimes called \textit{the stable unit treatment on neighborhood value assumption}~\cite{forastiere2020identification,leung2020treatment}. A larger $n$ implies a more relaxed assumption.
\item \textbf{(Specified network motifs)} Given the specified $n$-hop ego networks and causal network motifs, we assume that only the assignment conditions of specified network motifs affect the outcome of the ego observation. This assumption implies that we do not distinguish two ego networks with identical counts of specified network motifs but with different network structures. Considering more and higher-order would mitigate this issue.
\end{enumerate}

Ideally, we should consider a large $n$ and network motifs with more nodes because these are more relaxed assumptions. However, two issues may arise. First, it is typically computationally expensive to count network motifs of many nodes in an ego network of large $n$. There are many possible network motif patterns and potentially large counts. 
Second, related to the positivity requirement in the next section, we need all (or almost all) observations to contain all specified network motifs. Specification of too many network motifs may exclude a significant proportion of samples from the analysis. 

Note that although all our examples are in the undirected setting, it can be easily extended to directed networks (e.g., a directed edge from $i$ to $j$ indicates that $i$ reached $j$). In either undirected or directed networks, we should be cautious that the networks are pre-treatment  so that the network structure is not affected by the treatment assignments. For example, when we use a network where edges represent whether the two nodes have interactions, the interactions used should have happened before the treatment was assigned; otherwise this would bring about an issue known as post-treatment bias \cite{angrist2008mostly}. 

After counting causal network motifs for each ego node in our network, our next step is to convert the counts to features, which will be used in the next section. Let $\mathbf{X}_i$ denote an $m$-dimensional random vector, referred to as \textit{interference vector}. The interference vector has an important requirement: Each element of the random vector is \textit{``intervenable''} --- that is, the random treatment assignment affects the value of each element of the vector. The requirement  addresses the selection bias issue when we estimate the average potential outcomes. 

\begin{figure}
    \centering
    \includegraphics[width=0.8\linewidth]{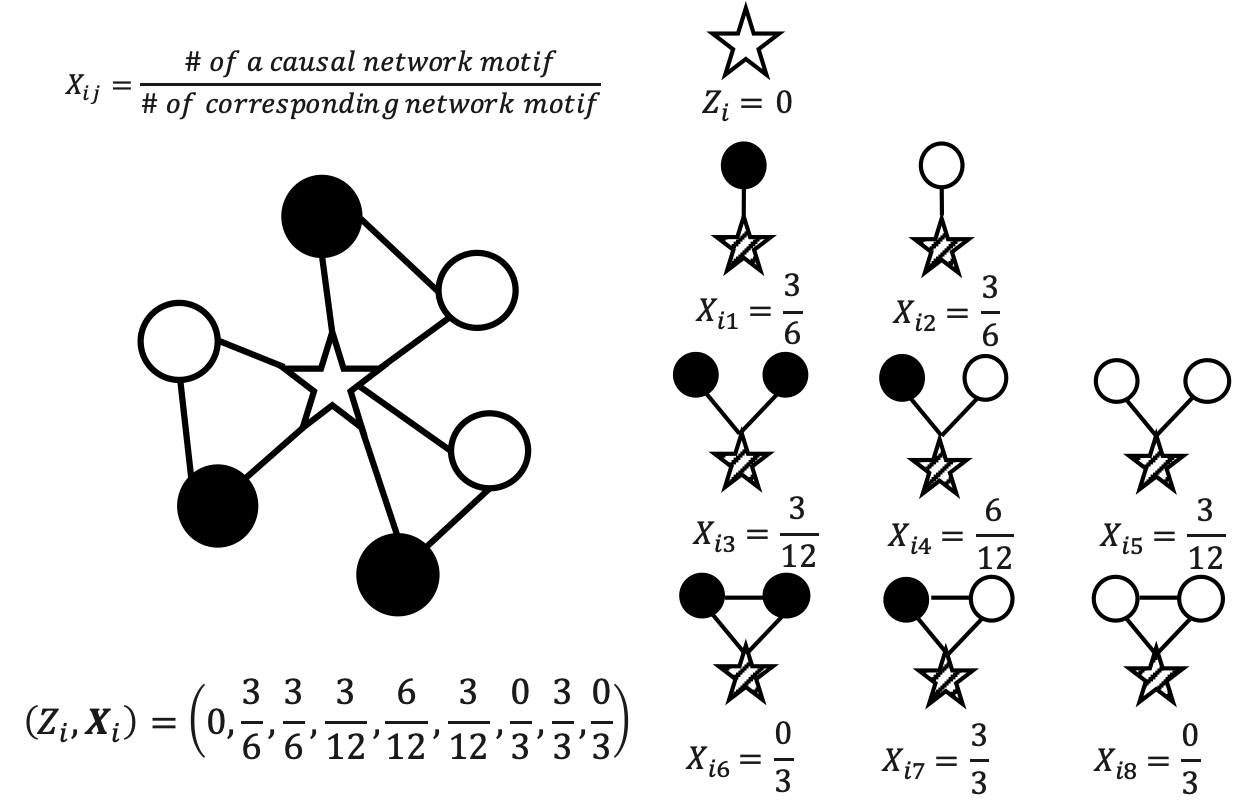}
    \caption{An example of ego network with treatment assignments and the corresponding interference vector. Stars represent egos and circles represent alters. Solid indicates the node being treated, hollow indicates control, and shaded indicates that it could be treated or control. }
    \label{fig:interference}
\end{figure}

We construct the interference vector in the following way. 
For each observation, for the count for each causal network motif (e.g., 2-1, 2-0, ..., 3o-2, 3o-1, ...), we normalize it by the count of the corresponding network motifs (e.g., dyads, open triads, closed triads, ...)\footnote{The $q$-fractional neighborhood conditions are considered special cases in our approach where only dyad motifs are used.}. In this way, each element of $\mathbf{X}_i$ is intervenable and the support for each element is in $[0, 1]$. Note that when considering a network motif with many nodes, some observations may not have certain network motifs, and normalization cannot be performed. In these scenarios, we can either exclude this network motif from the interference vector, or drop these observations if they take a really small proportion. Please refer to Figure~\ref{fig:interference} for an illustration of constructing the interference vector.

We combine the ego node's own assignment condition $Z_i$ and interference vector $\mathbf{X}_i$ as the features for our tree-based algorithm described in the next section to determine exposure conditions. $(Z_i, \mathbf{X}_i) \in [0, 1]^{m+1}$. Related to \cite{aronow2017estimating}, our approach is mathematically equivalent to, by an abuse of notation, $D=f(Z_i, \mathbf{X}_i)$ such that $\bar{y}(D) = \bar{y}(f(Z_i, \mathbf{X}_i))$. $\mathcal{D}$ is equivalent to partitioning $[0, 1]^{m+1}$ to $\mathcal{X}_1 \cup \mathcal{X}_2 \cup ... \cup \mathcal{X}_{|\mathcal{D}|}$. $Z_i$ impacts direct effects, and $\mathbf{X}_i$ corresponds to indirect effects.

\section{A Tree-Based Partitioning Approach}
\label{sec:tree}
\begin{figure}[t!]
    \centering
    \includegraphics[width=0.8\linewidth]{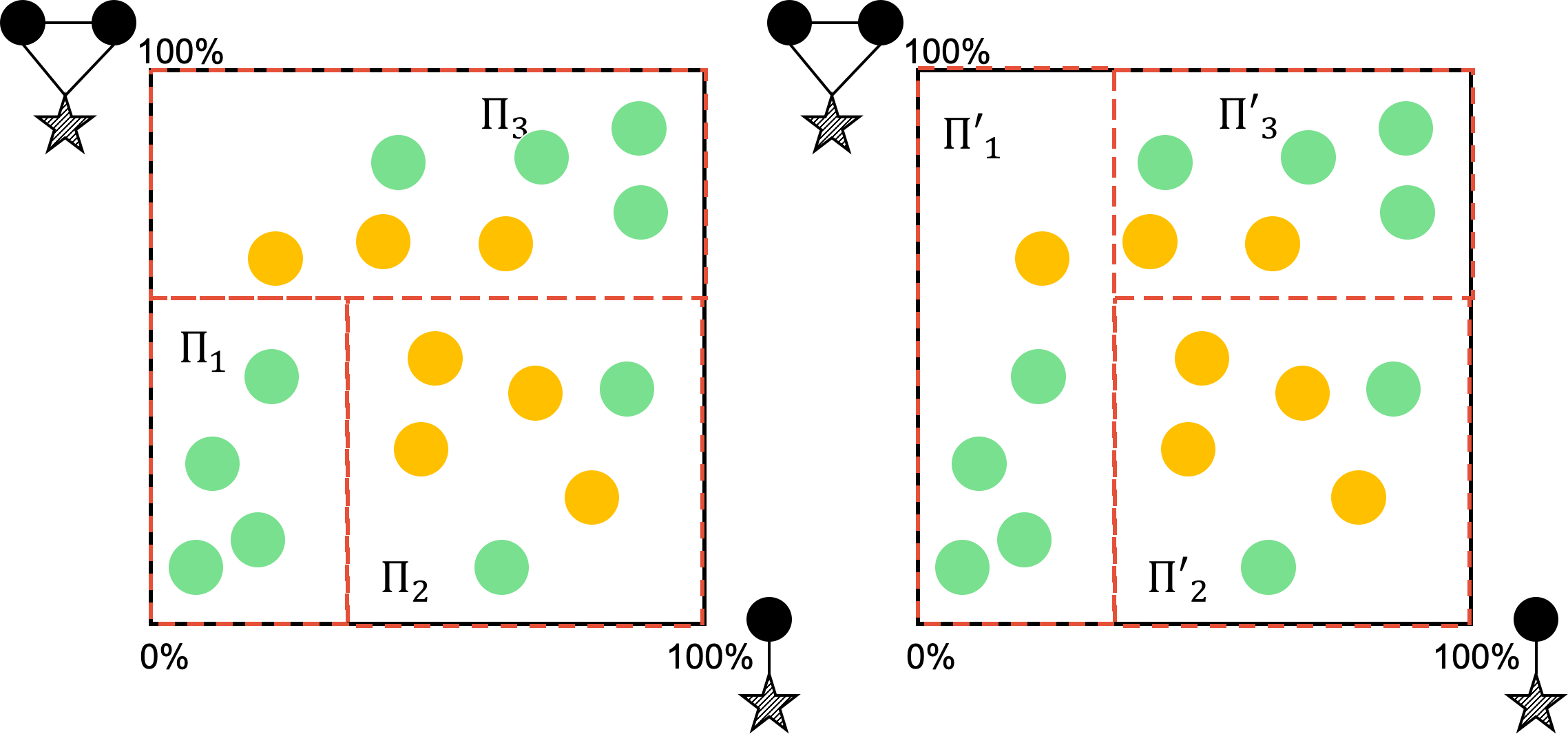}
    \caption{\small{\textbf{Illustration of selection bias and positivity}. $x$-axis and $y$-axis represent the fraction of the given causal network motif among the corresponding network motifs, assuming that we only specify these two causal network motifs. The positions of nodes indicate observed values for each observation (not the probability distribution).
    $\Pi_1$, $\Pi_2$, $\Pi_3$ or $\Pi_1'$, $\Pi_2'$, $\Pi_3'$ represent a plausible partitioning. Imagine green and yellow nodes represent two types of observations (e.g. green for observations with fewer neighbors and yellow for observations with more neighbors). In independent random assignments, green nodes are more likely to have extreme values in the $x$- or $y$- axis, while yellow nodes are more likely to be centered around the mean. 
    The left partitioning may violate positivity because yellow nodes may have zero or very small probability to belong in $\Pi_1$; by contrast, the right partitioning is feasible. 
    In the right partitioning, simply taking the average is still problematic because yellow nodes have a smaller probability to belong in $\Pi_i'$. Therefore, we need inverse probability weighting (e.g., Hajek estimator) to correct this selection bias.}}
    \label{fig:selection}
\end{figure}

The next question is how to design an algorithm to determine exposure conditions by $(Z_i, \mathbf{X}_i)$ (i.e., partitioning its support $[0, 1]^{m+1}$). By an abuse of notation, we replace any exposure condition $d$ by the corresponding partition $\mathcal{X}$ (where $\mathcal{X} \subset [0, 1]^{m+1}$): $\pi_i(\mathcal{X}) = \mathbbm{P} [(Z_i, \mathbf{X}_i) \in \mathcal{X}]=\mathbbm{P} [f(Z_i, \mathbf{X}_i) = d] =\pi_i(d)$, and similarly $1_i(\mathcal{X}) = \mathbbm{1} [(Z_i, \mathbf{X}_i) \in \mathcal{X}]=1_i(d) $, or the potential outcome $y_i(\mathcal{X})=y_i(d)$.

Our approach partitions $[0, 1]^{m+1}$ and determines exposure conditions based on a decision tree regression \cite{bishop2006pattern}.\footnote{We only illustrate regressions because they can be generalized to binary outcome variables, but our approach could be easily extended to classification.}  Decision trees can be used for clustering \cite{liu2000clustering} and typically have good interpretability in the decision making process \cite{quinlan1986induction}. Thus, it is a proper machine learning algorithm to solve the partitioning problem. Each leaf of the decision tree corresponds to  a unique exposure condition (partition). Compared with conventional decision tree regression, we need to have the following revisions:

\begin{enumerate}
    
    \item \textbf{(Positivity)} Positivity ensures observations have a non-zero chance of being in an exposure condition \cite{westreich2010invited}. 
    In our setting, it means that for all $i \in \mathcal{U}$, $d \in \mathcal{D}$, $\pi_i(\mathcal{X}) > 0$. 
    If any partition $d$ would lead to the existence of any observation $i$ that $\pi_i(d) = 0$, we would be unable to estimate the average potential outcomes. Mathematically, it would make the denominator in Equation~\ref{eq:hajek} zero. Note that the requirement is to set all observations ($\mathcal{U}$), rather than just the observations that are randomly assigned to $\mathcal{X}$.
    
    This is also part of the reason why we normalize the interference vector. Imagine we use the number of each causal network motif as the elements of the interference vector. If we decided a partition in which the number of 3c-2 (closed triads with fully treated neighbors) is greater than 10, then all observations with fewer than 10 open triads would have zero probability of belonging to this partition. The necessity of positivity is illustrated in Figure~\ref{fig:selection}.

    The tree algorithm should not split a node if such splitting will lead to any child node corresponding to an exposure condition in which any observation has zero probability to belong. Since $\pi_i(\mathcal{X})$ is sometimes not solvable analytically, we use Monte Carlo to approximate it ($\hat{\pi}_i(\mathcal{X})$) as mentioned previously for the estimation of ($\hat{\pi}_i(d)$).  Moreover, we adjust the positivity requirement to non-trivial probability, which allows a very few observations to have zero or near-zero probability; non-zero $\delta$ and $\epsilon$ introduce a small bias but allow partitioning more features. 

    \begin{equation}
        \sum_{i \in \mathcal{U}} \mathbbm{1} [\hat{\pi}_i(\mathcal{X}) \leq \epsilon ] \leq \delta |\mathcal{U}|.
        \label{eq:non-trivial}
    \end{equation}\noindent It means that the fraction of observations with $\hat{\pi}_i(\mathcal{X}) \leq \epsilon$ is smaller than $\delta$.

    \item \textbf{(Honest splitting)}  One pitfall is that when we estimate the variance, 
    the algorithm is choosing a threshold to partition so that it would minimize its objective function; however, it may overfit the training data, selecting an improper threshold when splitting, and eventually overestimating the difference between the average potential outcomes indicated by the two child nodes. We thus split the original training set into training and estimation sets --- the training set is used for tree partitioning and a separate estimation set is used for estimating the mean and variance. This is a common approach to correcting confidence intervals when using machine learning for causal inference \cite{athey2016recursive,kunzel2019metalearners}. 
    
    \item \textbf{(Weighted sum of squared errors)} Hajek estimator can be derived by minimizing  weighted sum of squared errors given a candidate exposure condition, which corresponds to a subset of $[0, 1]^{m+1}$ (denoted by $\mathcal{X}$).
    \begin{equation}
    \hat{\bar{y}}_{Hajek}(d) = \text{arg}\min_{y} \sum_{i \in \mathcal{U}, 1_i(\mathcal{X})=X} \frac{1}{ \pi_i(\mathcal{X}) } (Y_i - y)^2.
    \end{equation}

    In conventional decision tree regression, one common criterion that determines whether the algorithm will continue to split a node is the reduction in sum of squared errors. 
    Since Hajek estimators can be solved through weighted linear regressions, it is sensible to use a \textit{weighted sum of squared errors} (WSSE). It can be computed through a weighted linear regression where the weight is $1/\pi_i(\mathcal{X})$.\footnote{In practice, we replace it by $1/\hat{\pi_i}(\mathcal{X})$.}
    \begin{equation}
    \text{WSSE}(\mathcal{X}) = \sum_{i \in \mathcal{U}, 1_i(\mathcal{X})=1} \frac{1}{ \pi_i(\mathcal{X}) }  (Y_i - \hat{\bar{y}}_{Hajek}(\mathcal{X}))^2.
    \end{equation}

    When considering splitting the partition $\mathcal{X}$ to sub-partitions $\mathcal{X}_{l}$ and $\mathcal{X}_{r}$, 
    we cannot simply compare $\text{WSSE}(\mathcal{X}_{l}) + \text{WSSE}(\mathcal{X}_{r})$ versus $\text{WSSE}(\mathcal{X})$. 
    This is because $\mathbbm{E}[ \sum_{i \in \mathcal{U}, 1_i(\mathcal{X})=1} \frac{1}{ \pi_i(\mathcal{X})} ]$ $=$ $\mathbbm{E}[ \sum_{i \in \mathcal{U}, 1_i(\mathcal{X}_l)=1} \frac{1}{ \pi_i(\mathcal{X}_l)} ]$ $=$ $\mathbbm{E}[ \sum_{i \in \mathcal{U}, 1_i(\mathcal{X}_r)=1} \frac{1}{ \pi_i(\mathcal{X}_r)} ]$ $=$ $|\mathcal{U}|$, $\text{WSSE}(\mathcal{X}_{l}) + \text{WSSE}(\mathcal{X}_{r})$ is generally greater than $\text{WSSE}(\mathcal{X})$. We thus adjust the splitting criterion by taking a weighted average between $\text{WSSE}(\mathcal{X}_{l})$ and $\text{WSSE}(\mathcal{X}_{r})$:
\begin{equation}
\begin{split}
 (\mathcal{X}_l^*, \mathcal{X}_r^*)  = & \text{arg}\min_{X_l, X_r} \frac{\sum_i {1}_i(\mathcal{X}_{l})}{ \sum_i {1}_i(\mathcal{X})}  \text{WSSE}(\mathcal{X}_{l}) + \\
 & \frac{\sum_i {1}_i(\mathcal{X}_{r})}{ \sum_i {1}_i(\mathcal{X})} \text{WSSE}(\mathcal{X}_{r}),   \\
   \text{ where  } &  \frac{\sum_i {1}_i(\mathcal{X}_{l})}{ \sum_i {1}_i(\mathcal{X})} \text{WSSE}(\mathcal{X}_{l})+ \frac{\sum_i {1}_i(\mathcal{X}_{r})}{ \sum_i {1}_i(\mathcal{X})} \text{WSSE}(\mathcal{X}_{r})\\
  & < \text{WSSE}(\mathcal{X}) - \gamma \\
  \text{and } & \sum_i {1}_i(\mathcal{X}_{l}) \geq \kappa \text{ and } \sum_i {1}_i(\mathcal{X}_{r}) \geq \kappa.
\end{split}
\label{eq:where}
 \end{equation}

$\gamma$ is a hyper-parameter used to require non-trivial reduction in WSSE. Similar to conventional decision trees, we can also set the minimum leaf size ($\kappa$) to prevent the tree from growing unnecessarily deep.\footnote{An empirically effective alternative is to require $\sum_{i \in \mathcal{U}, 1_i(\mathcal{X}) = 1 }{\frac{1}{\hat{\pi}_i(\mathcal{X})}}$ to be very close to $|\mathcal{U}|$; e.g., $ (1 - \phi)|\mathcal{U}|\leq \sum_{i \in \mathcal{U}, 1_i(\mathcal{X}) = 1}{\frac{1}{\hat{\pi}_i(\mathcal{X})}} \leq (1 + \phi) |\mathcal{U}| $, where $\phi$ is a small number. 
This helps us to avoid partitions with a small number of observations, which may have a high degree of randomness in  $\sum_{i \in \mathcal{U}, 1_i(\mathcal{X}) = 1 } { \frac{1}{\hat{\pi}_i(\mathcal{X})}} $ and consequently, $\hat{\bar{y}}_{Hajek}(\mathcal{X}) $.  }
However, if the constraints cannot be satisfied, $(\mathcal{X}_l^*, \mathcal{X}_r^*)$ is an empty set and the algorithm would not further split $\mathcal{X}$.
\end{enumerate}

\begin{algorithm}[t!]
\caption{Implementation for the tree-based algorithm}
\label{algo:tree}
\begin{algorithmic}[1]
\small{
\Procedure{Split}{$\mathcal{X}$}
    \State $\mathcal{X}_{l}^*, \mathcal{X}_{r}^* = \emptyset$
    \State Compute $\hat{\bar{y}}(\mathcal{X})$ and its variance using estimation set
    \State Compute WSSE($\mathcal{X}$) using training set
    \State $\text{WSSE}^*(\mathcal{X}) = \text{WSSE}(\mathcal{X})$
    \For {$k \in [1, m+1]$}
        \For {$i \in \mathcal{U}$} \label{random}
            \State $\theta \gets \mathbf{X}_{ik}$
            \State $\mathcal{X}_{l} \gets \{j| \mathbf{X}_{jk} \leq \theta \And j \in \mathcal{X} \}$ 
            \State $\mathcal{X}_{r} \gets \{j| \mathbf{X}_{jk} > \theta  \And j \in \mathcal{X} \}$ 
            \If{Eq~\ref{eq:non-trivial} is true for both $\mathcal{X}_{l}$ and $\mathcal{X}_{r}$ }
                \State Compute WSSE($\mathcal{X}_{l}$), and WSSE($\mathcal{X}_{r}$) using training set
                \If {Constraints in Eq.~\ref{eq:where} are satisfied}
                    \State $\mathcal{X}_{l}^* \gets \mathcal{X}_{l}$
                    \State $\mathcal{X}_{r}^* \gets \mathcal{X}_{r}$
                    \State Update $\text{WSSE}^*(\mathcal{X})$ using Equation~\ref{eq:where}
                \EndIf
            \EndIf
        \EndFor
    \EndFor
    \If {$X_l^* \neq \emptyset$\text{ and }$X_r^* \neq \emptyset$} \Comment{If find any satisfy Eq.~\ref{eq:where}, choose the one with minimum WSSE.}
        \State \textsc{Split}($\mathcal{X}_l^*$)
        \State \textsc{Split}($\mathcal{X}_r^*$)
    \Else
        \State Add new exposure condition to $\mathcal{D}$, corresponding to $\mathcal{X}$
    \EndIf
     \EndProcedure
}
\end{algorithmic}
\end{algorithm}

Our algorithm is implemented by recursion (see Algorithm~\ref{algo:tree}). 
\textsc{Split} is a procedure used to partition a given space. 
One can use \textsc{Split}($[0, 1]^{m+1}$) to start the recursion algorithm. The data used for the algorithm is a random half of the original training set. We then use the separate half to estimate the mean and variance.  Empirically, direct effects (the impact of $Z_i$) are usually larger than indirect effects $\mathbf{X}_i$ (the impact of $\mathbf{X}_i$). 
Thus, the first partition is more inclined to split on $Z_i$: that is, it splits into two trees, in which one corresponds to treated observations and the other corresponds to non-treated observations. 

The bottleneck for computational efficiency is Line~\ref{random} --- the loop that iterates every element in $\mathcal{U}$. An improvement is to replace this for-loop by randomly choosing $\eta$ observations in $\mathcal{U}$. This may not help select the optimal cutoff but would provide good cutoffs if $\eta$ is not too small. We also have several hyperparameters --- $\gamma$, $\kappa$, $\delta$, $\epsilon$, and $\eta$. To tune those parameters, we can apply cross-validation to find the best choice that minimizes $\text{WSSE}^*([0, 1]^{m+1})$.


In addition to estimating the average potential outcome, our approach can also be applied to estimating global average treatment effects and heterogeneous direct effects with indirect effects fixed as explained: 
\begin{enumerate}
    \item \textbf{Global average treatment effects.} The global average treatment effect of an experiment is the difference between the average outcome in a counterfactual world where everyone is treated versus that in a counterfactual world where everyone is non-treated \cite{ugander2013graph}. This is fundamentally unsolvable unless assumptions such as SUTVA or our assumptions are made. In our approach, these two counterfactual worlds belong to two separate partitions. We can just identify these two partitions and compare the difference in the two average potential outcomes to compute the global average treatment effect. 
    
    Specifically, when the tree algorithm terminates, we have two special subsets of $[0, 1]^{m+1}$, denoted by $\mathcal{X}_0$ and $\mathcal{X}_1$ respectively. $\mathcal{X}_0$ is the partition that contains the cases where $Z_i=0$, all elements of $\mathbf{X}_i$ that represent fully non-treated neighborhood (2-0, 3o-0, 3c-0, 4o-0, ...) are equal to 1, and the rest equal 0; 
    $\mathcal{X}_1$ is the partition that contains the cases where $Z_i=1$ and all elements of $\mathbf{X}_i$ that represent fully treated neighborhood (2-1, 3o-2, 3c-2, 4o-2, ...) are equal to 1, and the rest equal 0. Then we can use $\hat{\bar{y}}(\mathcal{X}_1)-\hat{\bar{y}}(\mathcal{X}_0)$
    to estimate the global average treatment effect.
    Variance can be estimated by the method proposed in \cite{aronow2017estimating}.
    
    \item \textbf{Heterogeneous direct effects with indirect effects fixed.} As discussed, our algorithm tends to split on $Z_i$ first if direct effects are larger than indirect effects. Once the tree grows to two sub-trees that correspond to treated and non-treated observations, respectively, the further partitioning is not synchronized in these two trees. However, what if we want to understand how direct effects vary across different network interference conditions? Let $\mathcal{X}^{(1)}$ and $\mathcal{X}^{(0)}$ be the two spaces such that $Z_i^{(1)}=1$ for all $(Z_i^{(1)}, \mathbf{X}_i^{(1)}) \in \mathcal{X}^{(1)}$; and $Z_i^{(0)}=0$ for all $(Z_i^{(0)}, \mathbf{X}_i^{(0)}) \in \mathcal{X}^{(0)}$.  $\mathcal{X}^{(1)}$ and $\mathcal{X}^{(0)}$ are complementary if for all $(0, \mathbf{X}_i^{(0)}) \in \mathcal{X}_i^{(0)}$, we have $(1, \mathbf{X}_i^{(0)}) \in \mathcal{X}_i^{(1)}$, and  for all $(1, \mathbf{X}_i^{(1)}) \in \mathcal{X}_i^{(1)}$, we have $(0, \mathbf{X}_i^{(1)}) \in \mathcal{X}_i^{(0)}.$
    
    We can then revise our tree algorithm by only partitioning on a space $[0, 1]^{m}$ (i.e., on $\mathbf{X}_i$) and not partitioning on $Z_i$. In each node, including leaves, the goal is no longer estimating the average potential outcome. Instead, we estimate the heterogeneous direct effects ($\mathcal{\tilde{X}} = \{\mathbf{X} |(1, \mathbf{X}) \in \mathcal{X}^{(1)} \} $)
    \begin{equation}
        \tau_{direct}(\mathcal{\tilde{X}}) =  \bar{y}_{Hajek} (\mathcal{X}^{(1)}) - \bar{y}_{Hajek} (\mathcal{X}^{(0)})
    \end{equation}

    $\mathcal{X}^{(1)}$ indicates the cases where neighborhood assignment conditions are in a given region (e.g., high structural diversity, high echo chamber, or any other interference conditions) and
    the ego is treated; and $\mathcal{X}^{(0)}$ indicates the cases where the ego is non-treated but neighborhood interference conditions are the same. In this case, $\tau_{direct}(\mathcal{\tilde{X}})$ represents the average direct effect under that interference conditions. 
    
    If estimating heterogeneous direct effects with indirect effects fixed, two main revisions in Algorithm~\ref{algo:tree} are made. First, we  adjust the weighted linear regressions. Remember that the coefficient for the constant variable represented the Hajek estimator; in this case, we need to add the $Z_i$ into the regression, and report the coefficient for this variable to estimate $\tau_{direct}(\mathcal{\tilde{X}})$. WSSE is still derived from the error term for the weighted linear regression. The partitioning space becomes $[0,1]^m$. Second, the variance of the estimator is more complicated to estimate. Essentially, the variance estimation is similar to  the methods proposed in \cite{athey2016recursive}. 
    
\end{enumerate}

\section{Experiments}
\label{sec:experiment}

We evaluate our algorithm in a synthetic Watts-Strogatz (WS) network where we can verify our approach recovers the ground-truth which we know, and a real-world A/B test of a new product feature on Facebook to illustrate the scalability of our approach. 

\subsubsection*{Watts-Strogatz Simulation Network.} In studies of causal inference, a common challenge is that the ground-truth potential outcomes under either treatment or control for a given unit $i$ is missing --- that is, for each observation, we can only observe one single potential outcome given either a treatment or control assignment; with presence of network interference, we can only observe the outcome under one exposure condition. Therefore, we rely on a simulation study, which is often used to verify the effectiveness of causal inference methods.

We generate a Watts-Strogatz network \cite{watts1998collective} with $|\mathcal{U}|$=200,000. The Watts-Strogatz model is a random graph generator that preserves network properties such as clustering and the ``small-world phenomenon.'' To ensure a large variation in the local structure of individuals' neighborhood, we set a high rewiring rate for edges --- 50\%. We set the number of replicates $R=100$.

To be as general as possible and also complement the real-world independent assignment experiment, we use graph cluster randomization \cite{ugander2013graph} in the simulation study. We use a simple clustering approach --- we cluster every 10 nodes on the ring of the WS network, and we assign the treatment randomly and independently on the cluster level. We consider the following four different data-generating processes for outcomes, assuming that $\varepsilon_i$ is randomly drawn from Gaussian distribution with the mean of 0 and the variance of 1:

\begin{enumerate}
\item \textbf{Cutoff Outcome}. 
\begin{equation}
Y_i^{(1)} = 0.1 |\mathcal{N}_i| + gender_i + 2 Z_i \times \mathbbm{1}[X_{i, 3c-2} > 0.7] + \varepsilon_i.
\label{eq:y1}
\end{equation}

\noindent $gender_i$ is a  covariate independent of any other independent variables, which is randomly assigned to 1 or 0. Since this function has clear cutoffs, we can use it to validate that our tree-based algorithm splits on the corresponding cutoff. 
\item \textbf{Causal Structural Diversity Outcome}. 
\begin{equation}
\begin{split}
Y_i^{(2)} = 0.1 |\mathcal{N}_i| + gender_i + structural\_diversity\_of\_treated + \\ Z_i \times structural\_diversity\_of\_treated + \varepsilon_i
\end{split}
\end{equation}

\noindent Structural diversity is the number of disjointed components in an observation's neighborhood (or ego network). ``\textit{structural\_diversity\_of\_treated}'' is defined as the structural diversity given the set of treated neighbors. 
Although it was found that structural diversity predicts a higher product adoption rate~\cite{ugander2012structural}, it is unclear whether this is a causal impact or just correlation. A causal impact means it is actually the structural diversity of the treated neighbors that matters, while correlation means that the behavior is reflected by network structure, regardless of assignment conditions~\cite{su2020experimental}.
Our approach can analyze whether the causal impact of structural diversity exists in experimental data. Especially, the numbers of fully treated open triads or tetrads indicate the structure diversity of treated neighbors (3o-2 or 4o-3), and we expect our algorithm to split on these features. 
\item \textbf{Correlational Structural Diversity Outcome}. 
\begin{equation}
\begin{split}
Y_i^{(3)} = 0.1 |\mathcal{N}_i| + gender_i + structural\_diversity + \\ Z_i \times structural\_diversity +  \varepsilon_i    
\end{split}
\end{equation}

\noindent ``\textit{structural\_diversity}'' is the structural diversity given all neighbors, regardless of their assignment conditions. Therefore, we do not expect the tree algorithm to split on any features.
\item \textbf{Validation Under Null Effect}. 
\begin{equation}
Y_i^{(4)} = |\mathcal{N}_i| + \varepsilon_i.    
\end{equation}

As a sanity check, we use as a covariate -- the number of neighbors as the outcome variable. We do not expect the tree algorithm to split on any features. 
\end{enumerate}

\begin{figure*}[th!]
    \centering
    \includegraphics[width=0.9\linewidth]{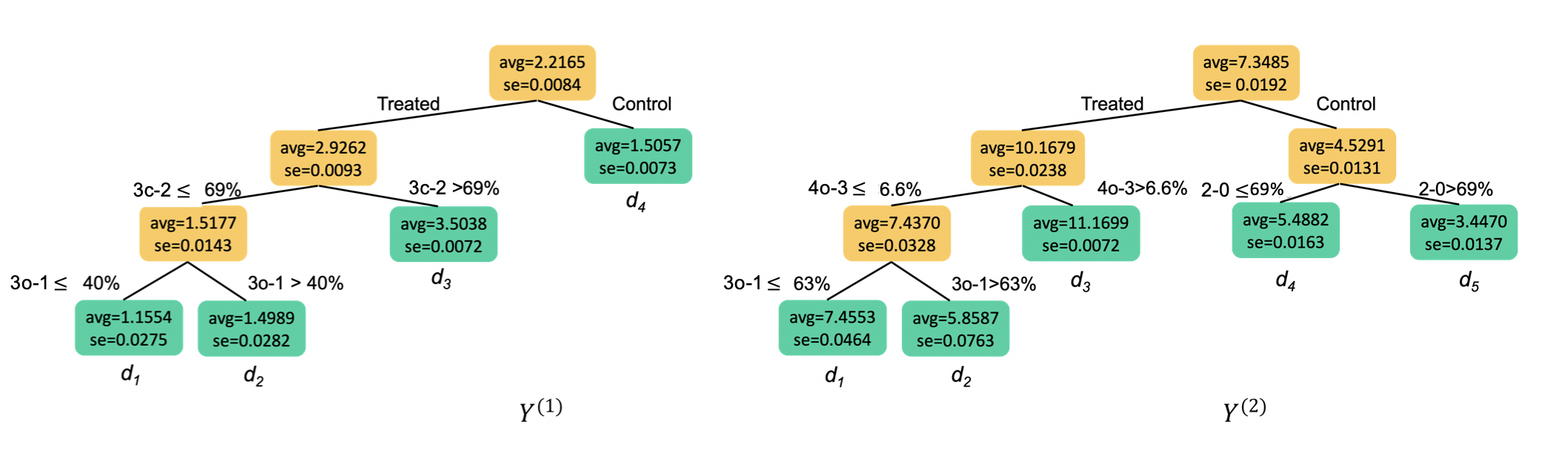}
    \caption{\small{\textbf{The result trees for the simulation experiment using all specified network motifs.} The two trees represent $Y^{(1)}$ or the Validation Cutoff Outcome (left) and $Y^{(2)}$ or the Causal Structural Diversity Outcome (right), respectively. The numbers in each leaf represents the average potential outcome and standard error (square root of variance) of the corresponding partition (exposure condition).}} 
    \label{fig:simulation}
\end{figure*}

\noindent Results are presented in Figure~\ref{fig:simulation} and summarized below:

\begin{enumerate}
    \item \textbf{Cutoff Outcome} ($Y_i^{(1)}$): As expected, after splitting treated versus control, it splits on fully treated closed triads (3c-2), with a threshold of 69\%. This is consistent with the parameter 0.7 set in Equation~\ref{eq:y1}. Also, as expected, it does not further split on control groups. However, the algorithm has an unexpected split on 3o-1, although the difference between the resulting $d_1$ and $d_2$ is not statistically significant. With further investigation, we find that this is because when partitioning the tree using the training set, it overfits the noise; however, since we use the estimation set for the average potential outcome and its variance (i.e.~honest splitting), the resulting tree does not show this significant difference. This result also demonstrates the importance of honest splitting. 
    
    \item \textbf{Causal Structural Diversity Outcome} ($Y^{(2)}$): The result is presented in the right panel in Figure~\ref{fig:simulation}. It first splits on treated versus control. For treated observations, it first splits on fully treated open tetrad (4o-3), which is positively correlated with the degree of structural diversity of treated neighbors. When the fraction of fully treated open tetrad is greater than 6.6\%, the algorithm terminates splitting, providing a partition (exposure condition) with the largest average potential outcome. If the fraction of fully treated open tetrad is smaller than 6.6\%, it further splits on open triads (3o-1). If the fraction of 3o-1 is greater than 63\%, the resulting exposure condition gives the smaller average potential outcome within treated observations. For control observations, partially because the structural diversity plays a smaller effect in the outcome function, it simply splits on the dyad level and terminates. 
    
    \item \textbf{Correlational Structural Diversity Outcome} ($Y^{(3)}$) and \textbf{Validation Under Null Effect}  ($Y^{(4)}$): As expected, it terminates after splitting samples into treated and control groups. Note these examples have nothing to illustrate.
\end{enumerate}

We compare our results with a baseline method used in \cite{aronow2017estimating}. The paper examines four exposure conditions: no effect, direct effect, indirect effect, and direct $+$ indirect effects. For example, when $Y_i^{(1)}$ is the outcome, most observations belong  to either ``indirect effect'' ($1.4981$ $\pm$ $0.0035$, 48.1\%) or ``direct $+$ indirect effects''  ($2.9182$ $\pm$ $0.0034$, 50.0\%). Similar results are derived in other outcomes. Therefore, this specification approach is not suitable for our experimental data. Moreover, such a partition cannot reveal the effects of specific network motifs or understand theories such as structural diversity. 


We also compare our result to the specification of the fractional $q$-neighborhood exposure condition in \cite{ugander2013graph}, which is equivalent to the proposed algorithm using dyad features only. This comparison can help highlight the importance of accounting for network motifs rather than simply counting treated friends.  When using only dyad features, the algorithm splits treated observations on 2-1 (i.e. fraction of treated neighbors), resulting in an average of potential outcome of $2.0472$ ($\pm 0.0189$) and $3.2137$ ($\pm 0.0085$) for less than or equal to, or greater than 60\%, respectively. It does not split the control observations either. 

\begin{figure}
    \centering
    \includegraphics[width=0.96\linewidth]{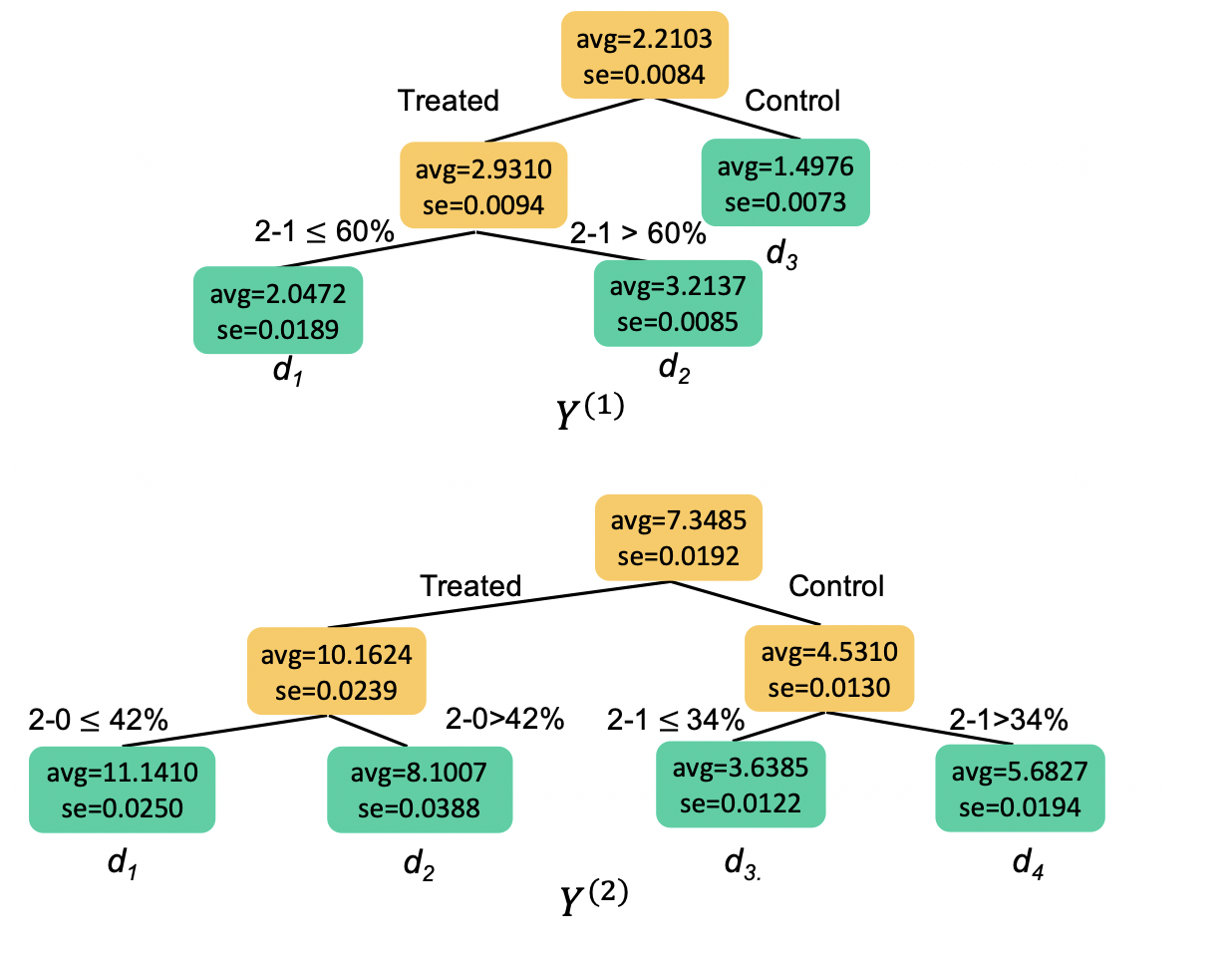}
    \caption{\small{The result trees for the simulation experiment using dyad network motifs only. The two trees represent $Y^{(1)}$ or the Cutoff Outcome (top) and $Y^{(2)}$ or the Causal Structural Diversity Outcome (bottom), respectively. The numbers in each leaf represents the average potential outcome and standard error (square root of variance) of the corresponding partition (exposure condition).}}
    \label{fig:bb}
\end{figure}

We use estimated global treatment effects \cite{ugander2013graph,chin2019regression} as a reference for the advantage of using causal network motif over simply accounting for proportions of treatment neighbors. Here we use $Y_i^{(1)}$ but similar results are derived in other settings. Using dyads features only (i.e., the proportion of treated neighbors, see Figure~\ref{fig:bb}) gives ($3.2137 - 1.4976 = $) $1.7161$ ($\pm0.0112$) while using all proposed motif features provides ($d_3$ and $d_4$ in the left panel of Figure~\ref{fig:simulation}, $3.5038 - 1.5057 = $) $1.9981$ ($\pm0.0102$); as a baseline,  the true global treatment effect is $2.0000$, and the average treatment effect under SUTVA is ($2.9262 - 1.5057 = $) $1.4205$ ($\pm0.0118$). 
In sum, using network motifs helps reduce more bias than using dyads features only when we estimate the global treatment effects because it characterizes network complex structure rather than purely counting the fraction of treated neighbors.

\begin{figure}
    \centering
    \includegraphics[width=0.55\linewidth]{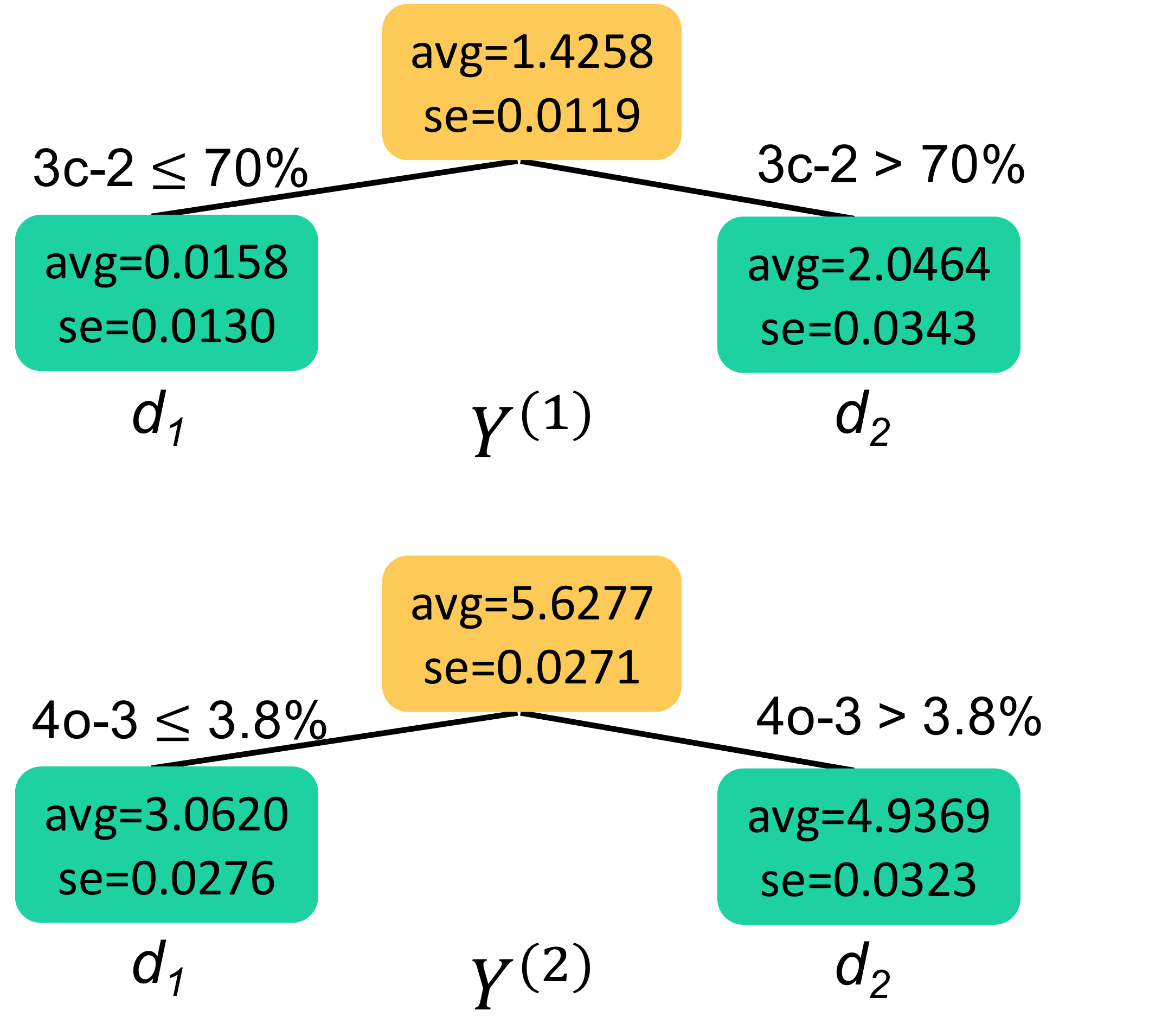}
    \caption{\small{\textbf{The result trees for heterogeneous direct effects for the simulation experiment using all specified network motifs.} The numbers in each node or leaf represents the average direct effect given the network interference condition and standard error (square root of variance) of the corresponding partition (exposure condition).}} 
    \label{fig:simulation_eht}
\end{figure}

We also investigate heterogeneous direct effects given different network interference conditions. As shown in Figure~\ref{fig:simulation_eht}, for both $Y^{(1)}$ and $Y^{(2)}$, it splits on the correct threshold: 
for $Y^{(1)}$, $Z_i$ only matters then the fraction of 3c-2 is greater than 70\%; for $Y^{(2)}$, it splits on important feature (4o-3), and the heterogeneous direct effect is larger when this fraction is greater.\footnote{Note that for the tasks of heterogeneous direct effects with indirect effects fixed, the average direct effect for a parent node does not necessarily lies between the values of its two child nodes. } As expected, no effect is observed for $Y^{(3)}$ and $Y^{(4)}$.

\subsubsection*{Care Reaction Rollout.} Finally, we apply our approach on a real-world A/B test of a product launch at Facebook.  Facebook launched the 7th reaction, ``Care''.\footnote{\small{\url{https://www.facebook.com/careers/life/the-story-of-facebooks-care-reaction}}}
When the reaction was launched, users in the control group (50\%) could not use the Care reaction but could still observe other users reacting with care. The treated group (the other  (50\%), could both use and see the Care reaction. 
The sample size is approximately 5\% of the Facebook population, and we only take into account users and neighbors who are in this 5\% sampled users. Note that we cannot compare treated users in this 5\% versus all the rest of Facebook population because the database does not log non-compliance: those who were assigned to the 5\% rollout but did not activate the treatment assignment (non-active users who did not attempt to use Care) are not recorded in the database and thus such a comparison is biased by non-compliance. 

\begin{figure*}[h]
    \centering
    \includegraphics[width=0.95\linewidth]{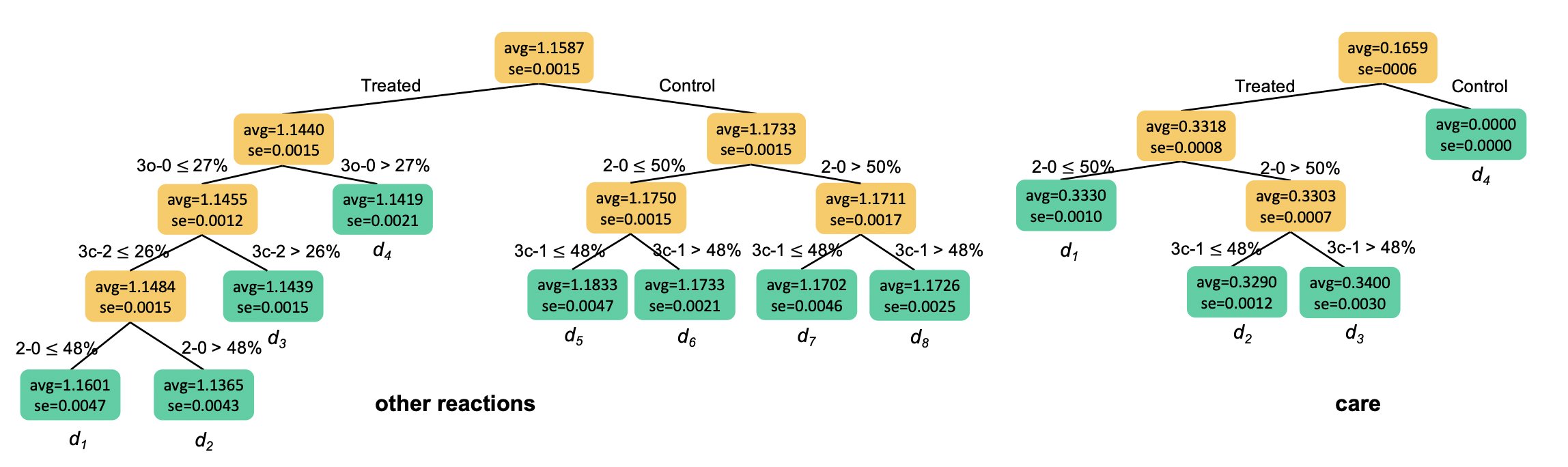}
    \caption{\small{The result trees for the Care experiment using all specified network motifs.} The left panel presents the result for the use of other reactions, and the right panel presents the result for the use of Care. The numbers in each leaf represents the average potential outcome and standard error (square root of variance) of the corresponding  partition (exposure condition).}
    \label{fig:care}
\end{figure*}

\begin{figure}[h]
    \centering
    \includegraphics[width=0.99\linewidth]{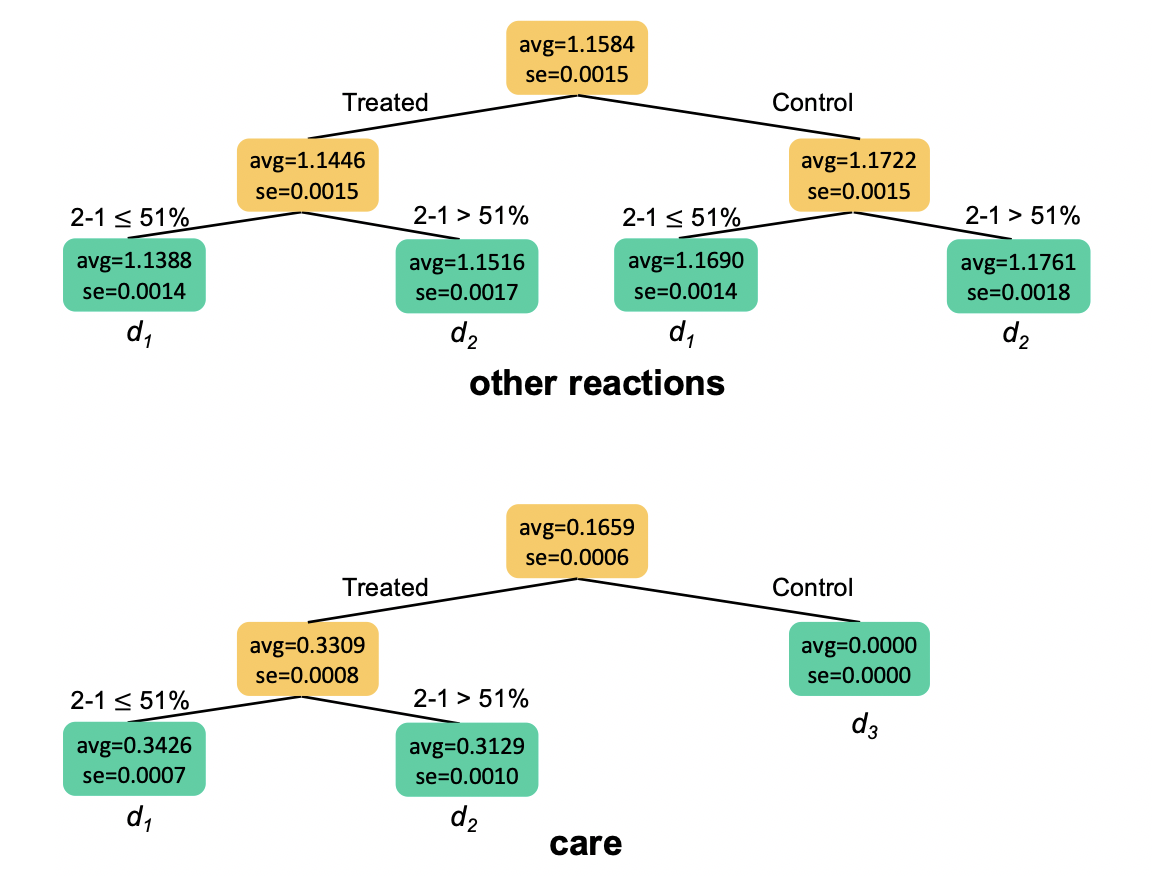}
    \caption{\small{\textbf{The result trees for the Care experiment using dyads only.} The upper panel presents the result for the use of other reactions, and the lower panel presents the result for the use of Care. The numbers in each leaf represents the average potential outcome and standard error (square root of variance) of the corresponding  partition.}}
    \label{fig:care_bb}
\end{figure}

In this experiment, we expect a user's use of reactions to be impacted by the number of their Facebook friends who could use Care and how those friends are connected which might impact their usage with each other. If a user has more friends using Care, or a group of friends able to use Care, she might use more Care reactions. 
There are two main outcome variables in this A/B test: (1) the number of Care reaction uses during the week after the experiment was launched. In the control group, the number of Care reaction uses is always zero; and (2) the number of other reaction uses. To prevent the impact from extreme points,  outcome variables are set on the log scale ($\log_{10}(x+1)$).

We randomly split our observations into approximately equal training
and estimation sets of sufficiently large sample sizes.
We use the training set to partition the tree and the estimation set for the average potential outcome and standard errors in each leaf node. 
We compute only dyad and triad network motifs, which is mainly restricted by computational resources.

The results are presented in Figure~\ref{fig:care}. There are eight exposure conditions determined by the algorithm when the outcome is other reactions, and four when outcome is use of Care. Again, we observe heterogeneity among different network interference conditions. For example, on the left panel, $d_1$ and $d_2$ have a difference of ($1.1601-1.1365 = $) $0.0236 \pm 0.0064$. As a reference, directly comparing the averages between treated and control groups gives $-0.0293$, which shows the importance of distinguishing between different network interference conditions.

Again, we also compare the results using all specified network motifs (Figure~\ref{fig:care}) versus the results using only dyads (Figure~\ref{fig:care_bb}) to illustrate the importance of using network motifs beyond dyads. 
In the left panel of Figure~\ref{fig:care}, $d_3$ contains the scenario of a fully treated neighborhood and $d_5$ the scenario of a fully non-treated neighborhood.  Therefore, we can estimate the global average treatment as $(1.1419-1.1833)=-0.0414$ ($\pm0.0049$). The baseline average treatment effect by directly comparing the averages between treated and control groups is $-0.0293$ ($\pm$0.0021). Thus, simply comparing the difference in means between treated and control groups could underestimate the true treatment effect by 25\%. 
By contrast, in the upper panel of Figure~\ref{fig:care_bb},  comparing $d_2$ and $d_3$ only gives $(1.1516-1.1690=)$ $-0.0174$ ($\pm0.0022$).
Similar analysis and conclusions also apply when the outcome is Care. 

We also examine heterogeneous direct effects under different network interference conditions. Since use of Care is always zero for the control group, this analysis is only meaningful when the outcome is  other reactions. The algorithm splits on 3o-2 at 23\% but does not provide significant differences between the two child nodes; we do not provide a figure here.

\section{Discussion}
\label{sec:discussion}
Network interference is much more complicated than simply being described as the ``indirect effect.'' To examine and analyze heterogeneity of indirect effects in experimental datasets, we provide a two-step solution. We first propose and employ the causal network motifs to characterize the network interference conditions, and then develop a tree-based algorithm for partitioning. Our tree-based algorithm is interpretable in terms of highlighting which exposure conditions are important for defining potential outcomes, addresses selection bias and positivity issues, and avoids incorrect standard error concerns via honest splitting.

Given the large number of researcher degrees of freedom in existing approaches for network interference such as choosing the threshold for an exposure condition, our approach provides a simple way to automatically specify exposure conditions. 
In this way, researchers no longer need to define exposure conditions a priori, and the exposure conditions generated by the algorithm are suitable for the given data and experiment. 
We believe that methodological innovation for addressing network interference concerns in A/B tests on networks will continue to be an important area for development, and accounting for network motifs with treatment assignment conditions provides a useful way to detect heterogeneous network interference effects. 

Our work also falls at the intersection of graph mining and causal inference. Graph mining methods are aimed at analyzing network properties such as the average number of connections, the number of communities, and more~\cite{chakrabarti2006graph,cook2006mining}. We focus on one type of graph property called a motif~\cite{milo2002network,ahmed2017graphlet,adami2011information,gallagher2008leveraging}. 
 However, we are not aware of any applications that have used labeled network motifs in causal inference applications, which is the focus of this work. 

Moreover, we  highlight our approach should not be confused with the Causal Tree and relevant methods~\cite{athey2016recursive,wager2018estimation}. The goal of our approach is to partition on the space of random assignments, while the Causal Tree and similar methods partition only on the covariate space. 
The critical difference in our setting is that in each partition of the treatment space, all observations should have a (almost) non-zero probability of belonging to that partition. In this way, we can construct certain unbiased estimators, such as Hajek, to estimate the average outcome under certain treatment regions and thus quantify a causal impact. 
By contrast, in Causal Tree and relevant methods, covariates are not intervened by the experiment and each observation has only probability of zero or one of belonging to each partition. 
Therefore, their methods are primarily aimed at partitioning the covariate space (i.e., to identify heterogeneous treatment effects for sub-groups in the sample) while our approach is to partition the treatment space (i.e, to identify the causal effects of different treatment or more specifically, exposure conditions).

Practitioners using our approach may obtain important insights. For example, they could understand how to utilize social contagion for product promotion when they have constraints on the number of promos. Researchers may identify important network interference conditions that are not theorized in certain experimental settings.

There remain many open questions or future directions based on our approach. First, we can incorporate covariates into the algorithm such as demographic features. 
One way is to also allow partitioning on the covariate space as well as the treatment space as in this work. However, once the algorithm splits on a covariate, 
all the descendants of that node only estimate the average potential outcome for the subsample that satisfies the criterion. The other way is  to incorporate covariates in the weighted linear regressions in our algorithm. This helps reduce the variance for the estimators and improve the precision when estimating average potential outcomes. We may also want to account for tie strength as well.

Second, we may consider alternative machine learning algorithms. Decision trees are not the only choice in our setting. 
We use a decision tree partially because it is an interpretable machine learning algorithm and does not involve functional form specification, except for assuming constant potential outcome.
Our tree-based algorithm can in fact be improved by the Hoeffding tree \cite{domingos2000mining}, which provides a streaming algorithm to perform the partitioning efficiently. 
Instead of using sample split, we may improve our methods by the conformal prediction theory \cite{shafer2008tutorial}. 
Moreover, we can imagine using nearest neighbor based algorithms and local regression instead to estimate the potential outcome given any point or region in $[0, 1]^{m+1}$. In addition, parametric methods can also be used when the goal is specific about estimating the potential outcome at certain points (e.g., fully treated neighbors) or estimating the global treatment effect \cite{chin2019regression}. 

Finally, our approach can be extended to any experimental data with multiple variables or continuous treatment variables, which are not necessarily only controlled experiments on social networks. While existing causal inference literature has primarily studied single binary or categorical treatments, fewer studies have approached continuous or multiple variables \cite{hirano2004propensity,imai2004causal}. Our approach provides a way to automatically convert multiple or continuous treatment variables to categorical treatments. It would be interesting to further investigate how machine learning can be applied to causal inference for continuous or multiple treatments as well as adapting this approach to observational causal inference settings. 
\newline

\noindent\textbf{Code Availability Statement: } \url{https://github.com/facebookresearch/CausalMotifs}\\

\noindent\textbf{Acknowledgements: } We thank Lada Adamic, Dean Eckles, Brian Karrer, and Steven Li for their helpful comments. We also thank the ``Fresh from the arXiv'' seminar participants at the University of Oxford and the MIT Conference on Digital Experimentation (CODE) participants for useful comments and feedback.

\bibliographystyle{ACM-Reference-Format}
{\linespread{0.95} \selectfont
\bibliography{reference}
}

\end{document}